\documentclass[prb,twocolumn]{revtex4}

\usepackage{graphicx}
\begin{document}

\title{Paramagnetic reentrance of the ac-screening: Evidence of vortex avalanches in Pb thin films}

\author{A.~V.~Silhanek}
\affiliation{Nanoscale Superconductivity and Magnetism Group, Laboratory for Solid State Physics and Magnetism
K. U. Leuven\\ Celestijnenlaan 200 D, B-3001 Leuven, Belgium}

\author{S.~Raedts}
\affiliation{Nanoscale Superconductivity and Magnetism Group, Laboratory for Solid State Physics and Magnetism
K. U. Leuven\\ Celestijnenlaan 200 D, B-3001 Leuven, Belgium}

\author{V.~V.~Moshchalkov}
\affiliation{Nanoscale Superconductivity and Magnetism Group, Laboratory for Solid State Physics and Magnetism
K. U. Leuven\\ Celestijnenlaan 200 D, B-3001 Leuven, Belgium}

\date{\today}

\begin{abstract}

We have studied the influence of a square array of pinning centers on the dynamics of vortex avalanches in Pb thin films by means of ac- and dc- magnetization measurements. Close to the superconducting transition $T_c$ the commensurability between the vortex lattice and the pinning array leads to the well known local increments of the critical current. As temperature $T$ decreases, matching features progressively fade out and eventually disappear. Further down in temperature vortex avalanches develop and dominate the magnetic response. These avalanches manifest themselves as jumps in the dc-magnetization and produce a lower ac-shielding giving rise to a paramagnetic reentrance in the ac-screening $\chi^\prime(T)$. Within the flux jump regime two subregimes can be identified. Close to the boundary where vortex avalanches develope, the field separation between consecutive jumps follow the periodicity of the pinning array and a field and temperature dependent screening is observed. In this regime, the response also depends on frequency $f$ in agreement with theoretical models for magnetothermal instabilities. At low enough temperatures and fields, the screening saturates to a constant value independent of $T$, $H$, and $f$, where jumps are randomly distributed. We have also found that vortex instabilities occupy a larger portion of the $H-T$ diagram in patterned samples than in films without nanoengineered pinning sites. Finally, we discuss the possible origin of the vortex avalanches and compare our results with previous experimental and theoretical studies.

\end{abstract}


\maketitle

\section{Introduction}

If a type-II superconductor is cooled down in a zero applied field (ZFC experiment) and subsequently an external
field $H$ larger than the first critical field $H_{c1}$ is applied, flux-bearing vortices enter through the sample's borders until they are captured by pinning centers. As a consequence, the system achieves an inhomogeneous
flux distribution with a higher density of vortices near the borders that progressively decreases toward the center of the sample. The spatial variation of the locally averaged field $\mathbf{B}(\mathbf{r})$ gives rise
to supercurrents $\mathbf{J}$ in the sample that, in the stationary state, accommodate to be exactly the critical current $J_{c}$ everywhere.  The resultant inhomogeneous flux distribution of this so-called \textit{critical state}, represents a self organized state that under small perturbations (like local temperature fluctuations $\delta T_i$) can lead to vortex avalanches in order to maintain $J=J_c(H,T)$ in that region. Because of the dissipation produced when flux lines move, each avalanche gives rise to a heat pulse and therefore a local increment of the temperature $\delta T_f$. If $\delta T_f < \delta T_i$ the critical state remains stable under these perturbations, otherwise avalanches of vortices draw the sample to a highly resistive state.\cite{chabanenko}

The way in which the vortex lattice reacts to the local overheating is crucial to determine the subsequent dynamics of the system. If the diffusivity of the magnetic flux (given by the resistivity $\rho$) is smaller than the thermal diffusion  coefficient (given by $\kappa/C$, where $\kappa$ the heat conductivity and $C$ is the heat capacity), the hot spot propagates in a frozen magnetic and current distribution (dynamic approximation). On the other hand, if thermal diffusivity is smaller than magnetic diffusivity there is not enough time to remove or distribute the heat produced by the vortex motion (adiabatic heating).\cite{mints-review} In both cases, the stability criterion for the critical state reads, 

\begin{eqnarray}
\label{eq1}
&&\frac{s^2}{\epsilon}\left|J_c\frac{dJ_c}{dT}\right| < 1 , 
\end{eqnarray}
where $s$ is the characteristic sample dimension, $\epsilon= C/\mu_0$ in the adiabatic approximation and $\epsilon=\kappa /\rho$ in the dynamic approximation.\cite{mints-review}

At high temperatures and fields, where the critical current is small, vortex avalanches are usually not seen. As temperature and field decrease, critical current increases and below a certain boundary $H^*(T)$ vortex avalanches develop. In principle this scenario should apply with minor differences for both bulk samples and thin films. However, recent magneto-optical images (MO) have shown that in thin films, with $H$ applied perpendicular to the plane of the film, the flux pattern in the sample exhibits a richer morphology than the smooth progressive flux penetration observed in the bulk. For this particular geometry, flux invasion occurs via dendritic structures which cover a substantial portion of the sample's area and grow very fast (i.e. under adiabatic conditions).\cite{barkov} The thermal origin of the observed instabilities has been recently demonstrated by strongly suppressing the dendrite instabilities as the thermal contact is improved.\cite{baziljevich}   

Typically, the observed sudden penetrations of the flux front into the sample are accompanied by sharp jumps in the dc-magnetization.\cite{chabanenko,duran,esquinazi,johansen} These jumps generate a noisy response which can undermine the technological applicability and perspectives of superconducting devices at low temperatures. However, recent promising experimental results have shown that this noise can be substantially reduced by introducing an array of pinning centers,\cite{vanacken,wordenweber,crisan} at expenses of increasing considerably the region in the $H-T$ diagram where flux jumps occur.\cite{hebert}

Additionally, in a recent work, Aranson {\it et al.}\cite{aranson} have theoretically predicted that the presence of a periodic modulation of the critical current would give rise to a growth of the branching process of the dendrites in comparison with a homogeneous pinning distribution. Evidence supporting this picture was reported by Vlasko-Vlasov {\it et al.}\cite{vlasko-vlasov} who analyzed the MO images of Nb films patterned with a square lattice of holes. The authors show that as the field is progressively ramped up, first the flux enters from the edges in stripes with boundaries along the principal axes of the pinning array and then new stripes jump between previously developed stripes. The width of these stripes involves several units cell of the pinning array.

On top of that, magnetization measurements performed on Nb\cite{terentiev} and Pb\cite{hebert} films with a square array of holes show that the field separation between consecutive jumps commensurate with the period of the underlying pinning array. This is a surprising result since matching features are typically seen only very close to $T_c$ where intrinsic pinning and self-field effects are not relevant. Clearly, the influence of the vortex pinning in the morphology and dynamics of the flux penetration on these kind of systems is an issue that has not been fully addressed so far and deserves further investigations.

In this work we study the flux-jump regime in Pb thin films with and without periodic pinning by means of ac- and dc- susceptibility measurements. In the samples with a square antidot array several regimes can be identified as a function of temperature. For temperatures $T \leq T_c$, matching features appear when the vortex lattice commensurates with the pinning array. As temperature decreases these effects progressively fade out. At a certain field-dependent temperature the shielding power of the sample is dramatically reduced and a reentrance in the ac-screening $\chi^{\prime}(T)$ is observed. This effect is accompanied by a substantial increase of the dissipation $\chi^{\prime \prime}$. We demonstrate that the observed reentrance is related to the appearance of flux jumps in the sample. Finally we discuss the origin of the quasi-periodic jumps in terms of a multiterrace critical state.    

\section{Experimental Details}

The experiments were conducted on Pb thin films with different pinning arrays. The dimensions and critical temperature for each sample are summarized in Table \ref{table}. In all patterned samples the square antidot array consists of square pinning sites with lateral dimension $b = 0.8~\mu$m and period $d=1.5~\mu$m which corresponds to a first matching field $H_1=9.2$ G. Simultaneously with each patterned film we deposited a plain reference film on SiO$_2$ substrate which allows us to perform a direct and reliable comparison in order to discriminate the effects of the pinning array. From the temperature dependence of the upper critical field $H_{c2}(T)$ we have estimated a superconducting coherence length $\xi(0) = 33 \pm 3$ nm for all samples.

\begin{table}[ht]
\centering \caption{Lateral dimensions (w$_1$ and w$_2$), thickness ($t$), and critical temperatures ($T_c$) for all the films studied. AD stems from square array of antidots and BH from square array of blind holes.}

\begin{tabular}[b]{l c c c c}
\\
Sample & w$_1$ (mm) & w$_2$ (mm) & t (nm) & $T_c$ (K)\\
     \hline
AD15 & 1.9 & 2.0 & 13.5 & 7.10 \\
AD65 & 2.3 & 2.5 & 65 & 7.21 \\
BH100 & 3.2 & 3.3 & 100 & 7.22 \\
\label{table}
\end{tabular}
\end{table}

The details of the sample preparation can be found in Ref.[\onlinecite{sophie-crete}]. Briefly, the predefined resist-dot patterns were prepared by electron-beam lithography in a polymethyl metacrylate/methyl metacrylate (PMMA/MMA) resist bilayer covering the SiO$_2$ substrate. A Ge(20~\AA)/Pb/Ge(200~\AA) film was then electron-beam evaporated onto this mask while keeping the substrate at liquid nitrogen temperature. Finally, the resist was removed in a lift-off procedure in warm acetone. The BH100 blind hole array was fabricated by depositing an additional 25 nm-thick Pb film on top of a 75 nm-thick Pb film with an array of antidots.

The ac-susceptibility measurements $\chi(H,T)=\chi^\prime+i\chi^{\prime \prime}$ were carried out in a commercial Quantum Design-PPMS device with drive field amplitudes $h$ ranging from 2 mOe to 10 Oe, and the frequency $f$ from 10 Hz to 15 kHz. The data were normalized to have a total step $\Delta \chi^{\prime} =$ 1, with $H=$ 0 at low temperatures and ac drives. This system provides a temperature stability better than $0.5$ mK which is crucial for measurements near the critical temperature. The dc-magnetization measurements were obtained using a QD-MPMS SQUID magnetometer equipped with a 5 T magnet.


\section{Results and Discussion}

\subsection{Dc-magnetization}
\begin{figure}[htb]
\centering
\includegraphics[angle=0,width=90mm]{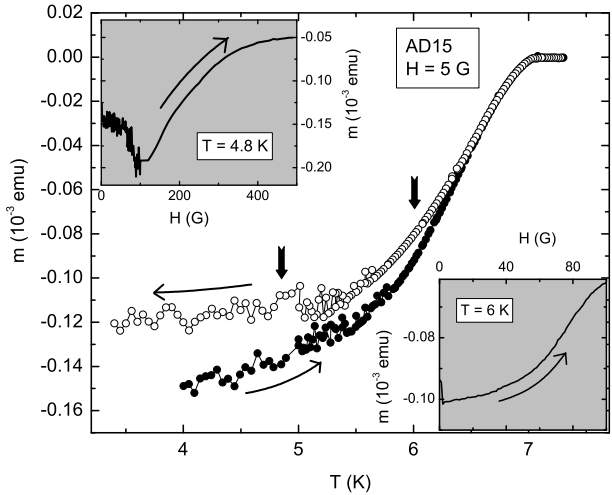}
\caption[]{{\small Main panel: Zero-field cooling (filled circles) and field-cooling (open circles) dc-magnetization as a function of temperature for the AD15 sample at $H=5$ G. The vertical black arrows indicate the temperature where the loops shown in the insets where recorded.}}
\label{fig1}
\end{figure}

In order to identify the temperature range where flux jumps occur, we have measured zero field cooled (ZFC) and field-cooled (FC) dc-magnetization for a fixed field $H$. The result of these measurements is shown in the main panel of Fig. \ref{fig1} for the AD15 sample at \mbox{$H = 5$ G}. A reversible response is obtained for $T_c-T < 0.7$ K, a temperature range higher than the expected for the irreversible line at this field,\cite{yoshida} probably due to the presence of an undesirable remanent field in the initial cooling procedure.

The AD15 sample is the same as that used in our previous studies.\cite{EJPB} In that work we show that commensurability effects between the flux line lattice and the pinning array manifest themselves as a peak in the critical current or as a higher screening in the ac-response. In this particular sample, well defined matching features appear only for $T_c-T \leq 0.6$ K. For lower temperatures (\mbox{$T<6.5$ K}), the decrease of the penetration depth $\lambda(T)$, the growth of the intrinsic pinning strength and self-field effects\cite{self-field} lead to a less pronounced matching features, which eventually disappear. Decreasing further the temperature there is a clear transition to a regime where magnetization becomes notably noisy. This transition occurs for both ZFC and FC curves around \mbox{$T = 5.7$ K} indicating that the onset of the crossover to a smooth behavior at high temperatures is highly reproducible. Of course the noisy response ($\delta m \sim 0.1 m$) is not related to the experimental resolution of the used device since in this regime the signal/noise ratio is higher than near $T_c$ where a smooth curve is obtained. Instead, we argue that at \mbox{$T = 5.7$ K} a transition to a more dissipative state owing to flux jumps takes place. Evidence corroborating this interpretation is obtained by measuring hysteresis loops at $T = 4.8$ K deep in the noisy regime (upper inset of Fig. \ref{fig1}) and at $T = 6$ K where the noise is absent (lower inset of Fig. \ref{fig1}). It can be seen that at \mbox{$T = 6$ K} jumps in the dc-magnetization never occur. In contrast to that, at $T = 4.8$ K flux jumps are present in the low field region up to $\sim 100$ G, where the average magnetization reaches a maximum. For \mbox{$H > 100$ G}, the irreversible magnetization decreases smoothly as field increases in agreement with a critical state scenario. As it has been pointed out previously, the magnetization peak at $100$ G indicates the onset of vortex avalanches in the sample.\cite{esquinazi}

\subsection{Ac-susceptibility}
\begin{figure}[htb]
\centering
\includegraphics[angle=0,width=90mm]{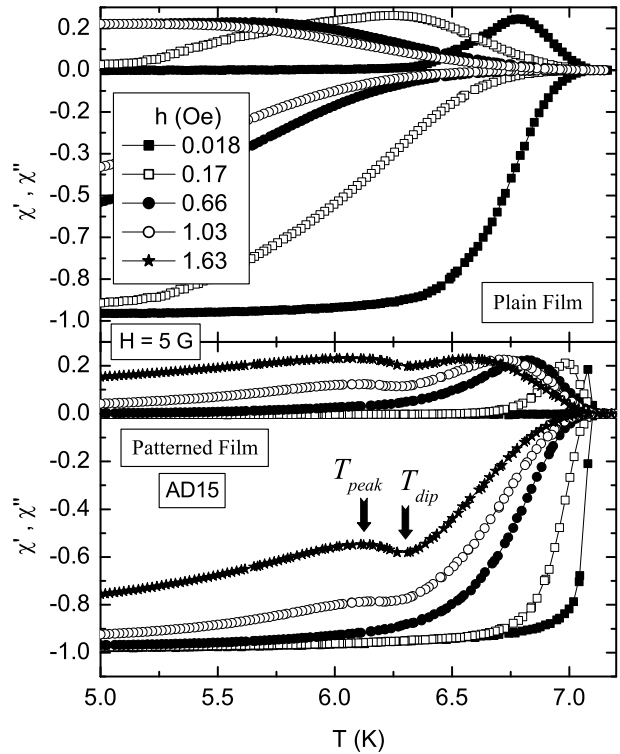}
\caption[]{{\small Temperature dependence of the ac-susceptibility $\chi=\chi^{\prime}+ i \chi^{\prime \prime}$ at several ac-drives $h$ for a plain film (upper panel) and a film with a square array of holes (lower panel). In the lower panel the transition to a flux-jumps regime appears as a reentrance signaled by the features $T_{dip}$ and $T_{peak}$ (see arrows).}}
\label{fig2}
\end{figure}

The appearance of flux-jumps should also be reflected as a lower efficiency to screen out an external ac field. This effect can be seen in the lower panel of Figure \ref{fig2} where the screening $\chi^{\prime}$ and the dissipation $\chi^{\prime \prime}$ as a function of temperature are shown for the AD15 sample at \mbox{$H=5$ G} and several ac excitations.

\begin{figure}[htb]
\centering
\includegraphics[angle=0,width=90mm]{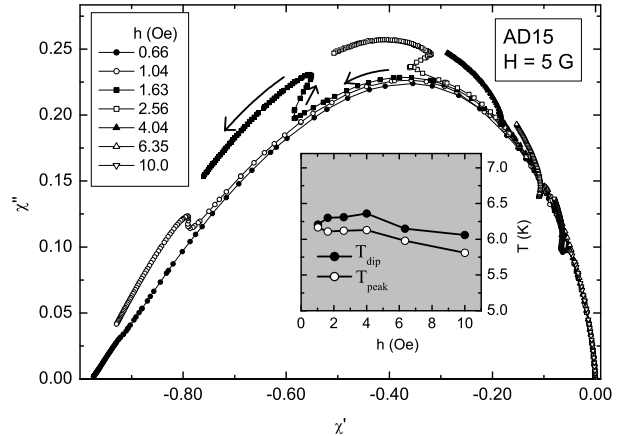}
\caption[]{{\small Main panel: $\chi^{\prime \prime}$ vs $\chi^{\prime}$ obtained from $\chi(T)$ curves at several ac-amplitudes $h$ for the AD15 sample (some of these curves are shown in the lower panel of Figure \ref{fig2}). The small arrows indicate the evolution of the curve as $T$ decreases. The inset shows the transition temperatures $T_{dip}$ and $T_{peak}$ (see Figure \ref{fig2}) as a function of $h$.}}
\label{fig3}
\end{figure}

At low ac drives no features indicating the transition to a flux-jumps regime are observed. For $h \geq 1$ G a paramagnetic reentrance in the screening at $T \sim 6.3$ K signals the onset of flux-jumps regime. It should be noted that the ac drive at which the reentrance in $\chi^\prime(T)$ first appears, $h \sim 1$ G, is the same order as the average distance between the jumps $H_j \sim 3$ G. This result suggests that as long as $h \ll H_j$ new vortex avalanches are not triggered and the transition to the flux jumps regime is not detectable. Consistently, we have observed a weak influence of the amplitude $h$ on the position where jumps first develop. This is shown in the inset of Figure \ref{fig3} where we can see that at low amplitudes, both the local minimum $T_{dip}$ and the local maximum $T_{peak}$, remain almost constant as $h$ increases. From $h = 4$ G up, the transition temperature slowly decreases with increasing $h$. As we will show below this effect is a consequence of a $T_{dip}$ decreasing with increasing the total field $H+h$. There is a systematic small discrepancy between the onset of the flux-jumps regime determined by dc-magnetization and ac-susceptibility. This effect can be attributed to a faster onset of vortex avalanches due to the ac shaking.

It is worth to note that the temperature evolution of the curves at temperatures above $T_{dip}$ and below $T_{peak}$ follows a quite different trend. Similar behavior has been recently reported by Passos {\it et al.}\cite{passos} for MgB$_2$ samples and attributed to a possible field induced granularity. The origin of the reentrant behavior in the screening can be understood from the upper inset of Fig.~\ref{fig1}. In this figure we can see that the development of avalanches leads to a reduction of the average critical current and therefore also to a reduction in the screening properties. In other words, when the system crosses from a ``non-jumpy'' regime to a ``jumpy'' regime, either by sweeping field or temperature, a strong suppression of the screening power occurs. As a rule, every peak in the dc-magnetization (either in field or temperature) will manifest itself as a dip in the ac-screening.\cite{giapintzakis,irie}

For comparison, in the upper panel of Figure \ref{fig2} we show similar measurements performed on a plain film (without nanoengineered pinning array) for the same ac drives and dc field. In the window of temperatures shown here, no features indicating the flux-jump transition are found. However, a reentrance appears at \mbox{$T_{dip} \sim 3.8$ K} for \mbox{$H=5$ G}. This reduction of $T_{dip}$ in the unpatterned sample is in agreement with recent magnetization measurements performed in samples with and without a periodic pinning array.\cite{hebert}

It should be noted that performing ac-susceptibility measurements the ``noise'' produced by the flux jumps is removed. This is so because the measurements here presented were obtained at a relatively high frequency $f=3837$ Hz and with an integration time of 1 sec, therefore the resultant $\chi$ reflect an average after cycling 3837 times a minor loop. By reducing the number of cycles one may approach to the case of dc-magnetization measurements and jumps become more apparent.

The transition to the flux-jump regime becomes more obvious by plotting the imaginary versus the real part of the ac-susceptibility: $\chi^{\prime \prime} \times \chi^{\prime}$ as shown in the main panel of Fig.~\ref{fig3}. The advantages of this so-called Cole-Cole plot is two fold, first it makes possible a comparison among different samples without knowing the specific value of the critical current,\cite{herzog} secondly it allows one to identify clearly different vortex dynamics regimes. In Figure \ref{fig3} we show this representation with $\chi$ obtained by sweeping temperature at several $h$ (part of these data were already shown in Figure \ref{fig2}). All these curves show a similar general behavior. Starting from high temperatures (right side) we first observe an increase in both dissipation and screening as $T$ decreases, until the transition temperature $T_{dip}$ is reached, then a sudden reduction of the screening together with an increase of the dissipation occurs down to $T_{peak}$. Below this temperature, the curves continue with a smooth evolution. Remarkably, the whole family of curves merges in two well distinguished evolvents, one for $T > T_{dip}$ and the other for $T < T_{peak}$, the latter being more dissipative than the former. Additionally, the peak of maximum dissipation is shifted from $\chi^\prime \approx -0.365$ for $T > T_{dip}$ to $\chi^\prime \approx -0.408$ for $T < T_{peak}$. This increase in magnitude of the maximum $\chi^{\prime \prime}$ together with the shift towards $\chi^\prime = -1$ is commonly observed in high temperature superconductors and attributed to creep effects. As we will discuss below the higher dissipative state is accompanied by a $f$-dependence susceptibility analogous to a creep regime where the frequency dependence of the magnetic response appears as a consequence of the dynamic evolution of the system (not considered in a conventional critical state model). 

\begin{figure}[htb]
\centering
\includegraphics[angle=0,width=90mm]{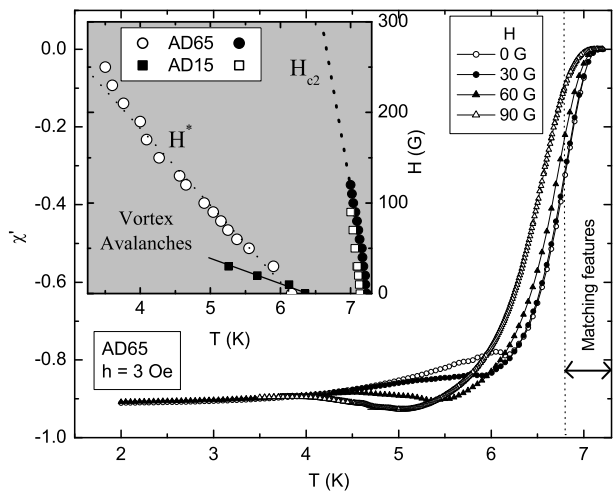}
\caption[]{{\small Main panel: $\chi^{\prime}(T)$ for the AD65 sample at $h=3$ G and several fields. The vertical dashed line indicates the lowest temperature where matching features are still barely defined. The inset shows the different observed regimes in a $H-T$ phase diagram for the AD15 (square symbols) and AD65 (circle symbols) samples. The boundary $H^*(T)$ represents the position of the local minimum in the $\chi^{\prime}(T)$ curves shown in the main panel.}}
\label{fig4}
\end{figure}

Once we have identified the temperature $T_{dip}$ (or $T_{peak}$) as the transition temperature to a flux-jumps regime, we can determine the portion of the $H-T$ diagram where these instabilities dominate by tracking the dip position as a function of the dc field. Some of these curves are shown in the main panel of Figure \ref{fig4} for the AD65 sample at $h=3$ G. In this figure it can be seen that the transition $T_{dip}$ shifts to lower temperatures as $H$ increases. This result is a consequence of a critical current $J_c$ decreasing with increasing $H$ and in agreement with previous experimental reports.\cite{hebert} It is also interesting to notice that at low temperatures and fields all these curves merge into a single universal behavior. This field independent curve indicates that the average irreversible magnetization in this region is also field independent. This is confirmed by the dc magnetization measurements shown in the main panel of Figure \ref{fig6} for the BH100 sample and by recent experiments performed on plain Pb films.\cite{radovan} Another important observation is that at low temperatures the universal magnetization seems to saturate to a value $\chi^\prime \approx -0.91$, and hence it also becomes temperature independent. Of course, at higher fields (\mbox{$H > 100$ G}) this field independent behavior is progressively lost as the system approaches the boundary $H^*(T)$.

This striking $T$- and $H$-independent average magnetization at low temperatures might be attributed to the coexistence of two different species of vortices. Indeed, it has been shown that in this regime dendrites have a finger-like structure with a size which remains unaltered with further changing the field. This very stable pattern of flux penetrated regions is a consequence of the tendency of dendrites to avoid each other. Once the field is reduced, already developed dendrites may help to remove flux from the sample and also channel incoming vortices when the field is again increased. In this picture, the ac response would be dominated by the easy motion of vortices inside the dendrites thus leading to a lower saturation value $\chi^\prime \approx -0.91$.

In the inset of Fig.~\ref{fig4} we show the resultant phase diagram obtained following the procedure described above for the AD65 (open circles) and AD15 (filled squares) samples together with the upper critical field $H_{c2}(T)$ for both samples.\cite{comment} An almost linear temperature dependence of the boundary $H_{dip}(T)$ is observed in the whole range of temperatures. Note that the observed different slopes of the boundary $H^*(T)$ (a factor of 3 steeper for the AD65 sample) cannot be ascribed to a different temperature dependence of $H_{c2}$ since the upper critical field is roughly sample independent. As a consequence, the $H-T$ region of flux-instabilities for the AD15 turns out to be smaller than for the AD65 sample. This reduction of the flux-instability region with decreasing the film's thickness is in agreement with previous observations in Nb thin films.\cite{esquinazi} Indeed, if $H$ is applied perpendicular to the sample's surface, the effective size $s$ which determines the stability criterion in eq.(\ref{eq1}) can be approximated as $s \sim \sqrt{wt/2}$, where $w$ is the width of the film.\cite{esquinazi,daumling} Using the sample dimensions shown in Table \ref{table} we obtain $s=3.8~\mu$m and $s=8.9~\mu$m for the AD15 and AD65 samples, respectively. From eq.(\ref{eq1}) we have that for dimensions $s$ such that $s^2 > s_{crit}^2=|\frac{1}{2}\epsilon \frac{dJ_c^2}{dT}|$ vortex instabilities appear. In other words, the smaller the $s$ the lower the boundary $H^*(T)$, in agreement with our observation. 

\begin{figure}[htb]
\centering
\includegraphics[angle=0,width=90mm]{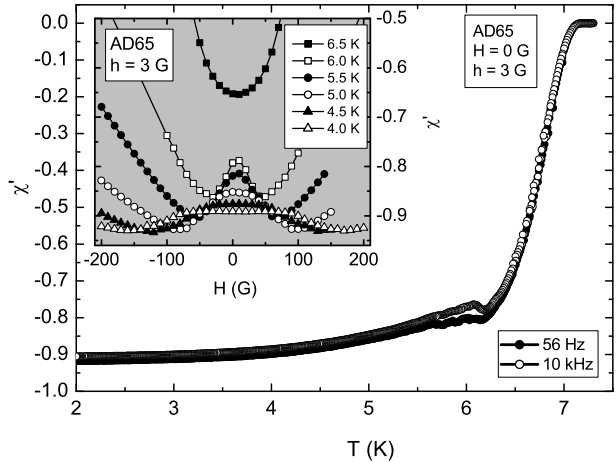}
\caption[]{{\small Main panel: $\chi^{\prime}(T)$ for the AD65 sample at zero field, $h=3$ G and two extreme frequencies. The inset shows the field dependence of the screening $\chi^{\prime}$ for several temperatures. At low $T$ and $H$, the flux-jump regime manifests itself as a reduction of the screening.}}
\label{fig5}
\end{figure}

As we pointed out previously, the transition to the vortex-instability regime manifests itself as a peak in the dc magnetization loops such as that shown in the upper inset of Fig.~\ref{fig1}. Similarly, we can determine this transition field by performing $\chi^\prime (H)$ measurements at several temperatures as is shown in the inset of Fig.~\ref{fig5} for the AD65 sample. In this figure, the data recorded at \mbox{$T=6.5$ K} fall, for all fields, in the stable regime where no jumps are observed and therefore exhibits the standard maximum screening at zero field. For the rest of the explored temperatures $T < 6$ K the maximum screening is no longer located at zero field but at $H_{peak}$ which roughly coincides with the boundary $H^*(T)$ shown in the inset of Fig.~\ref{fig4}. We observe that at low temperatures a plateau in $\chi^\prime (H)$ appears around $H=0$ in agreement with our previous remark. This constancy of $\chi^\prime(H)$ implies a fixed size of the flux avalanches and a field and temperature independent magnetization. This is consistent with recently reported avalanche distribution performed in similar samples\cite{hebert} and magnetization measurements on plain Pb films.\cite{radovan} According to Ref.[\onlinecite{hebert}], for \mbox{$T < 5$ K} the size distribution of flux-jumps does not depend strongly on temperature and in particular, the maximum jump remains almost constant. This effect is not inherent of samples with a periodic pinning array since similar results were found in unpatterned Pb films.\cite{radovan} At lower temperatures ($T < 3$ K) a peak in $\chi^\prime (H)$ around $H=0$ reappears although its maximum screening at $H=0$ is always smaller than $\chi^\prime=-0.91$. So, still this saturation value imposes an upper bound for the maximum possible screening in the flux-jump regime.

Lets now discuss what kind of instability corresponds to the observed behavior. An estimation of the ratio $\tau = \kappa \sigma / C$ between thermal and magnetic diffusivities, using\cite{jericho,novotny} $C=0.67$ J/Kg-K and $\kappa=3.4$ W/cm-K, and considering that $\rho \sim \rho_{ff}$ where $\rho_{ff}=\rho_n H/H_{c2}$ is the flux-flow resistivity and $\rho_n \sim 0.02 ~\mu \Omega$m  the normal state resistivity, gives $\tau \sim 1$ at $4$ K and $H=100$ G. Therefore, within this context neither the adiabatic nor the dynamic (or isothermal) approximations seems to be appropriate. However, in the estimate we made several important points are not included. First, the heat removal through the substrate which results in a lower local increase of the temperature.\cite{baziljevich} Second, the field and temperature dependence of the specific heat and the thermal conductivity. All these effects tend to reduce the value of the effective $\tau$ thus approaching to the adiabatic limit. In addition, our estimate is based on a critical state model where the dendritic penetration observed in these samples is not considered.\cite{vlasko-vlasov} Moreover, it is important to note that flux jumps are not mounted on a continuous prolongation of the curve observed at high field (see Fig.\ref{fig6}), thus suggesting that the observed vortex avalanches cannot be described within a critical state scenario. It is believed that in thin films the long-range vortex-vortex interaction and the non-local current-field relation are key ingredients which should eventually be incorporated in order to describe the branching process of the flux penetration.\cite{mints-brandt,barkov}

On the other hand, the similarity between the measurements presented in this work and those previously reported in several other materials (MgB$_2$,\cite{johansen} Nb,\cite{duran,vlasko-vlasov} Pb,\cite{radovan} Nb$_3$Sn\cite{rudnev}) points to magnetothermal instabilities as the origin of the flux jumps at low temperatures. As we mentioned in the Introduction, there is a rich zoology of possible vortex avalanches like smooth flux fronts, finger-like penetration or highly-branched tree-like flux invasion. In MgB$_2$, MO images indicate that quasi-unidimensional fingers occur at low temperatures whereas at intermediate temperatures a branched structure dominates.\cite{johansen} Similarly, Vlasko-Vlasov {\it et al.}\cite{vlasko-vlasov} showed that in Nb films patterned with an array of holes flux penetrates as stripes at low fields (resembling the fingers observed in MgB$_2$) whereas at higher fields a branching process produced by ``magnetic discharging'' connects the original stripes. This fast flux penetration strongly suggests that our results can be described within the adiabatic approximation. 

Another interesting result comes from the analysis of the frequency dependence of the screening. According to the commonly used theoretical models, in both limits, dynamic ($\tau \gg 1$) and adiabatic ($\tau \ll 1$), the size and the distance between jumps should decrease with increasing the field sweep rate $dH/dt$.\cite{mints-review,gerber} It has been also experimentally shown\cite{gerber} that the mean value of the magnetization, which is in fact what we are able to determine with ac-susceptibility, decreases with increasing $dH/dt$. In order to study this effect we have measured the $\chi^\prime(T)$ at $h=3$ G for two extreme frequencies \mbox{$f=56$ Hz} and $f=10$ kHz, as shown in the main panel of Fig.~\ref{fig5}. This frequencies correspond to a field sweeping rate of $0.05$ T/s an $12$ T/s, respectively, thus widely covering the range where $f$-dependence has been observed in other materials.\cite{gerber} Surprisingly, there is almost no frequency dependence unless in a narrow temperature window right below $T_{dip}$ where a smaller screening is detected for the higher frequency as expected.\cite{gerber} On the other hand, the lack of $dH/dt$ dependence in the low temperature regime has been also reported in Ref.[\onlinecite{esquinazi,hebert}] although in these publications the used $dH/dt$ covered a much smaller range. 

\subsection{Influence of the periodic pinning}

\begin{figure}[htb]
\centering
\includegraphics[angle=0,width=90mm]{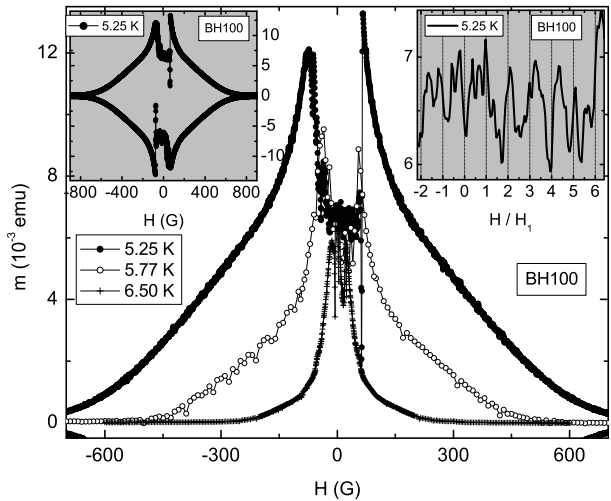}
\caption[]{{\small The main panel shows the decreasing branches of dc-magnetization loops recorded at three different temperatures for the BH100 sample. In the flux-jumps regime the average magnetization tends to a common value. A full hysteresys loop is shown in the left inset. The right inset shows a zoom-in of the flux jump regime where quasi-periodic jumps appear. The units used in the vertical axes of the insets is the same as the one used in the main panel.}}
\label{fig6}
\end{figure}

We now turn to the analysis of a possible matching between the periodicity of the jumps and that imposed by the pinning landscape. It has been recently shown, first by Terentiev {\it et al.}\cite{terentiev} in Nb films and later on by H\'ebert {\it et al.}\cite{hebert} in Pb films that within the flux-jump regime and at high enough temperatures and fields, the distance between consecutive jumps $H_j$ coincides with a multiple of the matching field $nH_1$, with $n$ integer. We have confirmed that this effect seems to be also present in the BH100 sample. Indeed, the upper right inset of Fig.~\ref{fig6} shows a zoom-in of the decreasing branch in the flux-jumps region for the BH100 sample at $T=5.25$ K. A Fourier spectrum analysis of these data shows that there are three maxima corresponding to $H_j=$ 12.8 G, \mbox{9.9 G} and 4.3 G which are relatively close to the matching conditions $H_j/H_1=$ 1.5, 1 and 1/2, respectively. Apparently these periodic jumps appear close to the boundary $H^*(T)$ with a jump size larger than that observed at lower temperatures and fields.

These {\it a priori} unexpected commensurability effects at low temperatures represent a striking observation since just above the transition $T_{dip}(H)$ there are no special features indicating the commensurability effects. The absence of matching features above $T_{dip}(H)$ is an indication that vortices are no longer regularly distributed. However, the influence of the periodic pinning array might still play an important role as it favors an easier motion of vortices along the rows of the array than at any other orientation.\cite{guided} This picture is consistent with the MO measurements of Ref.[\onlinecite{vlasko-vlasov}] showing preferential penetration along the principal axes of the pinning array. Moreover, recent MO images\cite{wijngaarden} performed on YBa$_2$Cu$_3$O$_{7-x}$ thin films with a square array of holes at very low temperatures ($T=4.5$ K) where {\it no matching effects are observed}, showed that the penetration of the flux front is highly anisotropic as a consequence of a strong guidance of vortices by the underlying pinning structure. 

Currently it is still a puzzling question why the period of the jumps coincides with multiples of $H_1$. In Ref.[\onlinecite{terentiev}] the authors postulate that at low temperatures, regular vortex structures may appear in a flux depleted region near the border of the sample as a result of geometric barriers (GB). As field is increased, this vortex-poor contour region is progressively filled with vortices which form a highly stable vortex pattern similar to those normally observed close to $T_c$. Above a certain threshold (i.e. $H_J=nH_1$) the vortex distribution in the border becomes unstable and is pushed towards the center of the sample by the screening currents thus triggering a flux avalanche. A first step in order to elucidate the origin of the observed quasi-periodic jumps is to discern whether GB are actually present in these samples.
    
Some evidence of the presence of GB comes from the fact that matching features obtained at high temperatures are systematically better resolved when decreasing field than increasing field as a consequence of a delayed entrance of vortices in the sample but without affecting their exit. However, in most cases the observed difference is almost imperceptible thus suggesting that the ubiquitous GB are not relevant. This result is consistent with the highly symmetric dc-magnetization loops observed {\it for all temperatures} studied as shown for example in the left upper inset of Fig.\ref{fig6}. In other words, if GB were indeed the responsible for the observed periodic jumps, this periodicity should be absent in the decreasing branch of the loop, in contrast to the general observation. Additionally, in a recent work we have shown that the ac-susceptibility response can be accurately described within the simple Bean critical state model for the perpendicular geometry without invoking the existence of GB. Indeed, in a system dominated by the GB, currents are constrained to flow along the sample's edge and therefore the sample's response could be modeled as that produced by a ring-shaped sample.\cite{herzog} This particular geometry will lead to a dome-like Cole-Cole plot more symmetric around $\chi^\prime=0.5$ than that shown for example in Fig.~\ref{fig3}. 

\begin{figure}[htb]
\centering
\includegraphics[angle=0,width=90mm]{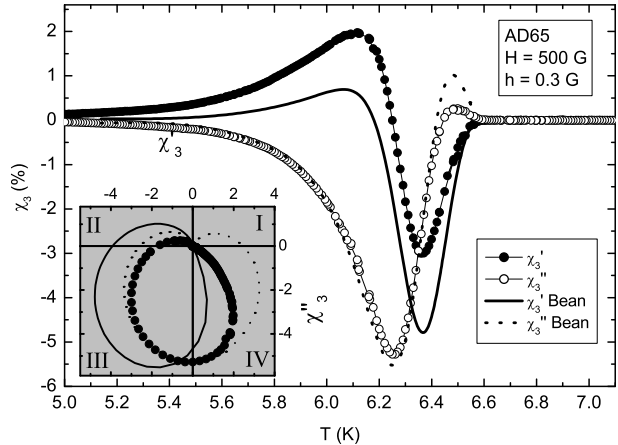}
\caption[]{{\small Main panel: Temperature dependence of the real (filled circles) and imaginary (open circles) components of the third harmonic susceptibility for the AD65 sample at \mbox{$H=500$ G} and $h=0.3$ G. The theoretical expectations for a disk in a critical state described by the Bean model are shown. In the inset the theoretical curve (solid line) and experimental data (filled circles) already shown in the main panel are represented in a Cole-Cole plot together with the curve for a ring shaped sample (dotted line).}}
\label{fig7}
\end{figure}

We can gain further insight in this particular issue by measuring the third harmonic susceptibility $\chi_3(T,H,h)$. The advantage of measuring this component lies in the asymmetry of the penetration-exit process which leads to a different shape of the minor loops traced out during an ac-cycle for the case of GB and bulk pinning.\cite{vanderbeek} Fig.~\ref{fig7} shows one of these measurements for the AD65 sample at $H =500$ G and \mbox{$h=0.3$ G}. Although we have also collected $\chi_3(T)$ data in the temperature range 5 K $<T<$ 7.5 K, at several dc fields (0 G $<H<$ 1 kG) and ac-drives (0.03 G $<h<$ 3 G), for all these conditions the curves are very much alike. In the same figure we have included the temperature dependence of the third harmonic components $\chi^{\prime}_3(T)$ and $\chi^{\prime \prime}_3(T)$ calculated using the Bean model for a disc shaped sample with field applied perpendicular to the plane of the disk.\cite{herzog,ishida} As we can see, the theoretical curves reproduce qualitatively the main features of the measured third harmonic. In contrast to that, these features cannot be accounted for by using the expression for a ring (similar to that obtained for edge barriers). This behavior becomes more evident in the inset of Fig.~\ref{fig7} where a Cole-Cole plot of the third harmonic using the same data shown in the main panel is presented together with the theoretical curve for the critical state in a disk shaped sample within the Bean model scenario. We also included in this figure the curve for a ring rescaled by a factor $\chi_0$ to fit with the experimental data and accounting for the expected amplitude dependence in the case of GB (strictly this is not valid but it rescues the basic heart-shaped curve obtained in the case of GB which allows one to make a fast comparison).

In the inset we clearly see that as for the disk geometry, the measured curve occupies the quadrants II, III and IV. In contrast to that, in the case of GB dominating behavior, the first quadrant (corresponding to low temperatures) should also be covered. This feature remains for all amplitudes and dc fields studied, thus suggesting that the contribution of
GB to the total response is very weak. On the other hand, the observed difference (mainly in  $\chi^{\prime}_3(T)$) between the Bean prediction and our experimental results, might be accounted for by assuming the more realistic case of field dependent critical current as in the Kim model.\cite{ozogul,maksimova}  

An alternative explanation for the observed periodicity in the flux jumps was recently proposed by H\'ebert {\it et al.}\cite{hebert}. According to these authors, at low temperatures the flux profile corresponds to a multiterrace critical state composed by steps of constant $B$ and zero critical current connected by abrupt changes in the flux density where the current is higher than that obtained from the average slope of the flux profile.\cite{cooley} In this scenario, the sudden penetration of a new terrace induces the movement of the internal terraces giving rise to jumps in the magnetization. At very low temperatures the concept of terraced state is lost as a more disordered flux distribution appears. On the other hand since the avalanches are triggered by the local slope rather than the average slope of the flux profile, this model satisfactorily accounts for the observed increase of the $H-T$ region where flux-jumps are detected in the patterned samples. However, a complete explanation of this puzzling observation remains elusive at the moment, and clearly further experimental and theoretical studies are needed.

In the final stage of preparation of this manuscript we learnt about a recent report by Zhukov et al.\cite{zhukov} showing several common results with the present work.

\section{Conclusion}

We have shown that the low temperature magnetic behavior of Pb films is dominated by vortex avalanches regardless the details of the pinning array. Although the size of the flux jumps is strongly reduced by introducing a periodic pinning array, the region of the $H-T$ plane where flux-jumps occurs is enlarged for patterned samples. This drawback can limit at low temeratures the applicability of these arrays to reduce the noise in SQUID systems as proposed recently.\cite{crisan} At low temperatures and fields an almost $T$- and $H$-independent magnetization is found. Although there is no clear clue to interpret this behavior, it might be related with the channeling of vortices by the predefined dendritic structures. Finally, we have demonstrated that the observed quasi-periodic jumps are unlikely to be originated in geometric barriers. Instead, a multiterrace critical state could satisfactorily account for this effect.

\acknowledgments
We would like to thank R. Jonckheere for fabrication of the resist patterns. This work was supported by the Belgian Interuniversity Attraction Poles (IUAP), Research Fund K.U.Leuven GOA/2004/02, the Fund for Scientific Research Flanders (FWO) and ESF ``VORTEX'' program.

\bibliographystyle{prsty}

\end{document}